\documentclass{article}

\usepackage[utf8]{inputenc}
\usepackage{amsmath}
\usepackage{amsfonts}
\usepackage{amssymb}
\usepackage{amsthm}

\usepackage{authblk}
\usepackage[width=13cm,height=20cm]{geometry}

\def\dif{\mathrm{d}}
\def\e{\mathrm{e}}
\def\iu{\mathrm{i}}

\def\SU{\mathrm{SU}}
\def\SO{\mathrm{SO}}
\def\nlsm{\mathrm{NL}\sigma\mathrm{M}}
\def\trans{\mathrm{T}}
\def\tr{\mathrm{tr}}

\begin{document}

\title{Chiral symmetry breaking and the Unruh effect \footnote{Contribution to the Fifteenth Marcel Grossmann Meeting. Rome, July 1-7. 2018.}}
\author{Adri\'{a}n Casado-Turri\'{o}n and Antonio Dobado}
\affil{Departamento de F\'{i}sica Te\'{o}rica and Instituto IPARCOS. Universidad Complutense, Madrid, 28040, Spain}


\maketitle

\begin{abstract}
The possibility of chiral symmetry restoration by acceleration is considered. The Thermalization Theorem formalism and the large $N$ limit (with $N$ being the number of pions) are employed to solve the lowest-order approximation to QCD at low energies in Rindler spacetime. It is shown that chiral symmetry is restored for accelerations higher than the critical value $a_c=4\pi f_\pi$, with $f_\pi$ being the pion decay constant. The results are completely analogous to those obtained in the inertial, finite-temperature case, evincing the ontic character of the Unruh effect.
\end{abstract}

\section{Introduction}

The decade of the 1970s saw the generalization of quantum field theory to curved spacetimes and arbitrary observers in Minkowski spacetime. This led to the establishment of several seminal results, such as the discovery by Hawking that black holes have an intrinsic temperature \cite{Hawking}, and thus radiate until they evaporate.

Attempting to better understand Hawking radiation, Unruh found \cite{Unruh} that a similar result may be obtained even when gravity is not present: he proved that a uniformly accelerated observer in Minkowski spacetime perceives the vacuum state of a free quantum field theory as a thermal ensemble of particles at temperature $T=a/2\pi$ (in natural units), with $a$ being the observer's acceleration. Despite the fact that this phenomenon requires experimentally unattainable accelerations to be observed directly (for instance, accelerations of the order of $10^{22}$ times the Earth's mean surface gravity are required to produce an Unruh temperature similar to ambient temperature), it is nevertheless considered to be highly fundamental, since it has been derived in several ways, such as the study of the response of particle detectors along non-inertial trajectories, canonical quantization \cite{FullingBirrelParkerBoulware} and even axiomatic quantum field theory \cite{Haag:1992hx} (for a thorough review of the Unruh effect, see Ref. \cite{Crispino:2007eb}).

However, none of these methods allows to discern whether the temperature--acceleration analogy introduced by the Unruh effect is just a formal result or if, on the contrary, it is capable of triggering non-trivial dynamical effects, such as phase transitions. The answer to this question requires the use of a formalism which is based on a path-integral approach to field theory, which will therefore be ideally suited to address arbitrary interacting theories and the connection between quantum field theory and thermodynamics. This formalism was developed by Lee \cite{Lee} in 1986, and is known as the Thermalization Theorem.

In this work, we shall make use of Lee's formalism in order to study whether accelerated observers are able to experience the spontaneous symmetry breaking or restoration of chiral symmetry, which is one of the main features of the fundamental theory of strong interactions, QCD, at finite temperature in Minkowski spacetime. Indeed, it has already been shown that accelerating observers detect the restoration of continuous global symmetries in some systems featuring Spontaneous Symmetry Breaking (SSB), such as the Nambu-Jona-Lasinio model \cite{Ohsaku:2004rv}, the $\lambda\Phi^4$ theory at the one-loop level  \cite{Castorina:2012yg} and the Linear Sigma Model ($\mathrm{L\sigma M}$) in the large $N$ limit \cite{Dobado}.

\section{Rindler spacetime and the Thermalization Theorem}

Consider a uniformly accelerated observer with proper acceleration $a$ along the $X$ direction of Minkowski spacetime $\dif s^2=\dif T^2-\dif X^2-\dif Y^2-\dif Z^2\equiv\dif T^2-\dif X^2-\dif X_\perp^2$, where $X^\mu=(T,X,Y,Z)$ are the usual Cartesian-like coordinates. The trajectory of such an observer is the hyperbola $T^2-X^2=-1/a^2$, whose asymptotes ---the null lines $T=\pm X$--- divide Minkowski spacetime into four quadrants: the regions $\pm X>|T|$, respectively known as the right ($R$) and left ($L$) Rindler wedges, each of which contains one of the branches of the hyperbola; and the regions $\pm T>|X|$, which are respectively the future ($F$) and past ($P$) of the origin. Notice that, while $ R$ and $L$ are causally disconnected, both share a common past, and so correlations between them may exist. Hyperbolic motion is conveniently described using comoving coordinates $x^\mu=(t,x,y,z)$, which range in $(-\infty,\infty)$ and are defined by
\begin{equation}
T=a^{-1}\e^{ax}\sinh(at),\hspace{10pt}X=\pm a^{-1}\e^{ax}\cosh(at),\hspace{10pt}Y=y,\hspace{10pt}Z=z,
\end{equation}
with the plus-sign choice covering only $R$ and the minus-sign choice covering only $L$. It is important to notice that this implies that the Euclidean Minkowski time $T_E\equiv\iu T$ and $X$ are periodic functions of the Euclidean comoving time $t_E\equiv\iu t$, with period $2\pi/a$. In terms of comoving coordinates, the Minkowski metric is
\begin{equation}
\dif s^2=\e^{2ax}(\dif t^2-\dif x^2)-\dif x_\perp ^2
\end{equation}
in both $R$ and $L$. The worldline with $x=0$ corresponds to the trajectory of constant proper acceleration $a$; in fact, each of the worldlines with constant $x$ is a trajectory with proper acceleration $a(x)=a\e^{-ax}$. Hence, it is also useful to define the so-called Rindler coordinates $\rho\equiv1/a(x)=\e^{ax}/a\in[0,\infty)$ and $\eta\equiv at\in(-\infty,\infty)$.

Let us suppose now that there exists a certain quantum field theory defined over Minkowski spacetime, which describes fields of arbitrary spin and their possible interactions. Consider, without loss of generality, a Rindler observer in $R$ (the results generalize straightforwardly to $L$). Because this observer is causally disconnected from all the events in $L$, she will be insensitive to any vacuum fluctuations outside $R$. As a result, the Minkowski vacuum state of the theory, which is a pure state $|\Omega_M\rangle$ for any inertial observer, becomes a mixed state in the eyes of the accelerated observer, who needs to perform a partial trace over the degrees of freedom in $L$. Most importantly, the corresponding density matrix was shown by Lee to be
\begin{equation}
\rho_R=\tr_L|\Omega_M\rangle\langle\Omega_M|=\dfrac{\e^{-2\pi H_R/a}}{\tr\,\e^{-2\pi H_R/a}},
\end{equation}
where $H_R$ is the Rindler Hamiltonian (i.e. the generarator of $t$-translations). This density matrix describes a thermal ensemble at temperature $T=a/2\pi$, which is precisely the Unruh temperature originally found for free scalar field theories. However, the applicability of this result is virtually universal: it guarantees that the Minkowski vacuum state will be perceived by a Rindler observer as a thermal state, independently of the field theory considered. This is the reason why it is also known as the Thermalization Theorem.

\section{Chiral symmetry and its restoration by acceleration}

Two-flavor massless QCD in ordinary Minkowski spacetime is known to possess two distinct phases, one in which chiral $\SU(2)_L\times\SU(2)_R$ symmetry is spontaneously broken into isospin symmetry $\SU(2)_{L+R}$, and one in which it is restored. Both phases are separated by a typical Landau-Ginzburg (i.e. second order) phase transition at a certain critical temperature $T_c$, with the order parameter being the quark condensate $\langle\bar{q}q\rangle_T$, where $q$ is the quark field and $\langle\,\cdot\,\rangle_T$ denotes the expectation value at temperature $T$ on the Minkowski QCD vacuum state. Thus, because of the temperature--acceleration analogy, it is natural to wonder whether a Rindler observer perceives a restoration of chiral symmetry if her acceleration $a$ is higher than a certain critical value $a_c$. This issue may be elucidated by functionally quantizing the low-energy effective theory for QCD\footnote{As it is well known, QCD is asymptotically free, and thus becomes non-perturbative at low energies. In this regime, quarks and gluons cease to be the relevant degrees of freedom and an effective description in terms of the next lightest particles in the hadronic spectrum, the pions, is needed. Coincidentally, the pions are also the (pseudo-)Nambu-Goldstone bosons associated to chiral symmetry breaking.} in the right Rindler wedge, and then computing the expectation $\langle\bar{q}q\rangle_a$ on the Minkowski QCD vacuum (which, as per the Thermalization Theorem, will be felt by the accelerated observer as a thermal ensemble at temperature $a/2\pi$). The lowest-order effective field theory for QCD is a non-linear sigma model ($\nlsm$) whose  Euclidean partition function in $R$ is
\begin{equation} \label{nlsmZ}
Z=\int[\dif\Phi][\dif\lambda]\exp\bigg(-\int\dif^4 x\,\sqrt{g}\,\bigg(\dfrac{1}{2}\partial_\mu\Phi^\trans\partial_\mu\Phi+\dfrac{\lambda}{2}(\Phi^\trans\Phi-f_\pi^2)-M_\pi^2f_\pi\sigma\bigg)\bigg),
\end{equation}
where $\sqrt{g}$ is the determinant of the Euclidean Rindler metric, $M_\pi$ is the pion mass, $f_\pi$ is the pion decay constant, and $\Phi^\trans=(\pi^a,\sigma)$ is a real quadruplet belonging to the fundamental representation of $\SO(4)\simeq\SU(2)_L\times\SU(2)_R$, consisting of the three pion fields $\pi^a$ (with $a=1,2,3$) plus an additional field $\sigma$ related to them through the non-linear constraint of the model, $\Phi^\trans\Phi=\pi^a\pi^a+\sigma^2=f_\pi^2$, which is in turn enforced through the introduction of the non-dynamical Lagrange-multiplier field $\lambda$. The fields are also required to satisfy the thermal-like Rindler boundary conditions $\pi^a(\tau=0,\mathbf{x})=\pi^a(\tau=2\pi/a,\mathbf{x})$, $\sigma(\tau=0,\mathbf{x})=\sigma(\tau=2\pi/a,\mathbf{x})$, $\sigma(\tau,|\mathbf{x}|=\infty)=f_\pi$ and $\lambda(\tau=0,\mathbf{x})=\lambda(\tau=2\pi/a,\mathbf{x})$. In addition, by comparing \eqref{nlsmZ} with the standard QCD partition function (restricted to $R$) it is clear that $\langle\bar{q}q\rangle_a\propto\langle\sigma\rangle_a$.

The $\nlsm$ model described above may be solved non-perturbatively in the large $N$ limit, where $N$ is the number of pions. This limit is properly defined if $f_\pi^2$ is of order $N$, i.e. $f_\pi^2\equiv NF^2$, with $F^2$ being $N$-independent. Bearing this in mind, the Euclidean partition function may be rewritten in the chiral limit $M_\pi\rightarrow 0$ as
\begin{equation} \label{PF5}
Z=\int[\dif\pi][\dif\sigma][\dif\lambda]\exp\bigg(-\int\dif^4 x\,\sqrt{g}\,\bigg(-\dfrac{1}{2}\pi^a\square\pi^a-\dfrac{1}{2}\sigma\square\sigma+\dfrac{\lambda}{2}(\pi^2+\sigma^2-f_\pi^2)\bigg)\bigg).
\end{equation}
The functional integral over the $N$ pion fields is Gaussian, and thus may be readily performed. Defining $\Gamma[\sigma,\lambda]$ as the effective action in the exponent of the remaining integral, the fields can be expanded around some point $(\bar{\sigma},\bar{\lambda})$ in the functional space where the first functional derivative of $\Gamma[\sigma,\lambda]$ vanishes. Using the steepest descent method, we have $Z=\e^{-\Gamma[\bar{\sigma},\bar{\lambda}]}+\mathcal{O}(N^{-1/2})$ and $\bar{\sigma}^2(x)=\langle\sigma(x)\rangle_a^2=\langle\sigma^2(x)\rangle_a$ in the large $N$ limit. Thus $\bar{\sigma}(x)$ and $\bar{\lambda}(x)$ are chosen to be the solutions of
\begin{eqnarray}
\dfrac{\delta \Gamma}{\delta \sigma(x)} & = & -\square\sigma+\lambda\sigma=0, \label{Equ1p} \\
\dfrac{\delta \Gamma}{\delta \lambda(x)} &  =  & \dfrac{1}{2}(\sigma^2 -f_\pi^2)+\dfrac{N}{2}G(x,x;\lambda)=0, \label{Equ2p}
\end{eqnarray}
with boundary conditions $\bar{\sigma}=f_\pi$ and $\bar{\lambda}=0$ at $x\rightarrow\infty$, and where
\begin{equation}
(-\square+\lambda)_x\,G(x,x';\lambda)=\dfrac{1}{\sqrt{g}}\delta^4(x-x').
\end{equation}

Unfortunately, an exact solution of these equations cannot be easily found. However, they may be solved approximately for $ax<<1$ (i.e. close to the origin of the accelerating frame) by exploiting the peculiar properties of Rindler spacetime. In this region, it can be shown \cite{Higgs} that $\lambda\simeq 0$, and that the relevant Green function is
\begin{equation} \label{Equ2p2}
G(x,x;0)=\int _0^\infty\dif\Omega\,\dfrac{\Omega\pi}{2\rho^2 \tanh(\Omega\pi)}.
\end{equation}
Introducing $\omega\equiv a\Omega$ and expanding $a\rho=1+ax+\ldots$, equation \eqref{Equ2p} becomes
\begin{equation} \label{reduEq}
\sigma^2=f_\pi^2 -\dfrac{N}{4\pi^2}(1-2ax)\int _0^\infty\dif\omega\,\omega\,\bigg(1 + \dfrac{2}{e^{2\pi\omega/a}-1}\bigg)+\mathcal{O}(a^2x^2).
\end{equation}
While the second integral on this equation may be easily calculated, the first one clearly requires regularization (for example, by introducing an $x$-dependent ultraviolet cutoff $\Lambda\e^{-ax}$, which leads to a renormalization of $f_\pi$). Thus, if we now define the $N$-independent critical acceleration as
$a_c^2\equiv 48\pi^2 f_\pi^2/N$
we have 
\begin{equation} \label{linear}
\bar{\sigma}^2(x)=f_\pi^2\bigg(1-\dfrac{a^2}{a_c^2}+2x\dfrac{a^3}{a_c^2}+\ldots\bigg),
\end{equation}
which is also a solution of equation \eqref{Equ1p} at this order. Assuming $\bar{\sigma}(0)$ to be real, the evaluation of this expression at the origin $x=0$ of the accelerating frame  yields
\begin{equation} \label{qqa}
\dfrac{\bar{\sigma}(0)\big|_a}{\bar{\sigma}(0)\big|_0}=\dfrac{\langle\sigma(0)\rangle_a}{\langle\sigma(0)\rangle_0}=\dfrac{\langle\bar{q}(0)q(0)\rangle_a}{\langle\bar{q}(0)q(0)\rangle_0}=\begin{cases}
\sqrt{1-\dfrac{a^2}{a_c^2}} & \text{if }0\leq a<a_c, \\
0 &  \text{if }a\geq a_c.
\end{cases}
\end{equation}
Thus, we clearly see that acceleration triggers the second-order chiral phase transition at the origin of the Rindler frame. For $N=3$ pions (i.e. the physical case for two-flavour QCD), the critical acceleration $a_c=4\pi f_\pi$ corresponds to an Unruh-like temperature $T_c=2f_\pi$, which is precisely the temperature at which the QCD phase transition is found to take place if the $\nlsm$ is solved in Minkowski spacetime \cite{Cortes:2016ecy}  in the same large $N$ limit that we have consider in $R$. Also, the same behavior for $\langle\bar{q}q\rangle_T/\langle\bar{q}q\rangle_0$ is found in this case, but substituting $a/a_c\mapsto T/T_c$. However, the expression for $\langle\bar{q}q\rangle_T/\langle\bar{q}q\rangle_0$ is valid in all Minkowski space, due to its homogeneity and isotropy.

We can extend our discussion to other points close to the origin by considering two different accelerated observers at Rindler coordinates $\rho=1/a$ and $\rho'=1/a'$. From the point of view of the first observer, the second one is located at some point with $x$ coordinate given by $\rho'=1/a'=\e^{ax}/a$, i.e. her acceleration is $a'=a\e^{-ax}$. Thus, the position-dependent condensate is given by
\begin{equation} \label{result}
\dfrac{\langle\bar{q}(x)q(x)\rangle}{\langle\bar{q}(0)q(0)\rangle}=\sqrt{1-\dfrac{a^2}{a_c^2}\e^{-2 a x}}.
\end{equation}
Therefore, for a Rindler observer with acceleration $a\in(0,a_c)$, the condensate ranges from $\langle\bar{q}(0)q(0)\rangle$ at infinity to zero at the critical value $x_c\equiv\ln(a/a_c)/a<0$. 

This means that the chiral phase transition takes place on the $(x_c,x_\perp)$ surface. The symmetry is also restored on the region close to the horizon $x< x_c$, where $\langle\bar{q}(x) q(x) \rangle=0$. Thus, the boundary between the broken and restored phases is completely determined by $a_c$, which only depends on QCD parameters.

Finally, it is interesting to notice that \eqref{result} implies that, for points different to the origin, what the condensate feels is equivalent to a thermal bath with a space-dependent temperature \cite{Candelas:1977zza} $T(x)\equiv a\e^{-ax}/2\pi$, which diverges at the horizon and goes to zero at infinity. This temperature is consistent with the Tolman and Ehrenfest rule \cite{Tolman:1930ona} for thermal equilibrium in static spacetimes, since $T(x)\sqrt{g_{00}}=a/2\pi$ is an $x$-independent constant, as required by the rule.

\section{Conclusions}

The powerful Thermalization Theorem formalism has allowed us to demonstrate the ability of the Unruh effect to produce non-trivial dynamical effects. In particular, we have been able to study the restoration of the highly fundamental QCD chiral symmetry by acceleration. We have shown that analogous results are found in both the thermal and Rindler cases, with a typical second-order phase transition occurring for those accelerated observers whose accelerations are higher than the critical value $a_c=4\pi f_\pi\simeq 1.6$ GeV for $N=3$ pions. The behavior obtained for the spacetime-dependent temperature felt by the condensate is also compatible with the standard requirements for thermodynamic equilibrium in static spacetimes, of which Rindler space is a particular example. Our results may have applications in ultra-relativistic heavy collisions, in which the Unruh effect has been proposed as a possible thermalization mechanism \cite{Kharzeev}, and also in black holes or cosmological scenarios, since local Rindler coordinates can always be defined close to any horizon \cite{Padmanabhan}.

\section*{Acknowledgments}
A. D. thanks Luis \'Alvarez-Gaum\'e  and C. Pajares for very interesting comments and discussions. Work supported by the Spanish grant  FPA2016-75654-C2-1-P.


\begin{thebibliography}{00}

\bibitem{Hawking}
  S.W. Hawking,
  \textit{Nature} \textbf{248}, 30 (1974);
  \textit{Comm. Math. Phys.} \textbf{43}, 199 (1975);
  \textit{Phys. Rev.} \textbf{D14}, 2460 (1976).
  
\bibitem{Unruh}
  W.G. Unruh,
  \textit{Phys. Rev.} \textbf{D14}, 870 (1976).
  
\bibitem{FullingBirrelParkerBoulware}
  S. Fulling,
  \textit{Phys. Rev.} \textbf{D7}, 2850 (1973):
  D.G. Boulware,
  \textit{Phys. Rev.} \textbf{D11}, 1404 (1975);
  \textit{Phys. Rev.} \textbf{D13}, 2169 (1976):
  L. Parker,
  \textit{Phys. Rev.} \textbf{D12}, 1519 (1976):
  N.D. Birrell and P.C.W. Davies,
  \textit{Quantum fields in curved space} (Cambridge University Press, 1982).
  
\bibitem{Haag:1992hx} 
  R. Haag,
  Berlin, Germany: Springer (1992) 356 p. (Texts and Monographs in Physics)
  
\bibitem{Crispino:2007eb} 
  L. C. B. Crispino, A. Higuchi and G. E. A. Matsas,
  \textit{Rev. Mod. Phys.} \textbf{80}, 787 (2008).
  
\bibitem{Lee} 
  T. D. Lee,
  \textit{Nucl. Phys.} \textbf{B264}, 437 (1986). 
  R. Friedberg, T. D. Lee and Y. Pang,
  \textit{Nucl. Phys.} \textbf{B276}, 549 (1986).
  
\bibitem{Ohsaku:2004rv} 
  T.~Ohsaku,
  Phys.\ Lett.\ B {\bf 599}, 102 (2004):
  D.~Ebert and V.~C.~Zhukovsky,
  Phys.\ Lett.\ B {\bf 645} (2007) 267.
  
\bibitem{Castorina:2012yg}
  P.~Castorina and M.~Finocchiaro,
  J.\ Mod.\ Phys.\  {\bf 3} (2012) 1703.
  
\bibitem{Dobado} 
  A.~Dobado,
  ``Spontaneous symmetry breaking and the Unruh effect,'' in {\it General Relativity, 1916-2016}
 (Minkowski Institute Press, Montreal 2017). arXiv:1703.05675 [gr-qc].

\bibitem{Higgs}
  A.~Dobado,
 ``Brout-Englert-Higgs mechanism for accelerating observers,''
  Phys.\ Rev.\ D {\bf 96} (2017) no.8,  085009
  
\bibitem{Cortes:2016ecy} 
  S. Cort\'{e}s, \'{A}. G\'{o}mez Nicola and J. Morales,
  \textit{Phys. Rev.} \textbf{D94}, no. 11, 116008 (2016).

\bibitem{Candelas:1977zza} 
  P.~Candelas and D.~Deutsch,
  \textit{Proc. Roy. Soc. Lond.} \textbf{A354}, 79 (1977).

\bibitem{Tolman:1930ona}
  R. Tolman and P. Ehrenfest,
  \textit{Phys. Rev.} \textbf{36}, no. 12, 1791 (1930).
  
\bibitem{Kharzeev}  
  D. Kharzeev, K. Tuchin,
  \textit{Nucl. Phys.} \textbf{A753}, 316-334 (2005).
  
\bibitem{Padmanabhan}
  T. Padmanabhan,
  \textit{Phys. Rept.} \textbf{406}, 49-125 (2005);
  \textit{Rep. Prog. Phys.} \textbf{73}, 046901 (2010).  

    


\end{thebibliography}
\end{document}